\documentclass[twocolumn,prl,superscriptaddress]{revtex4-1}

\usepackage{color,amsfonts,graphicx,verbatim}
\usepackage{amsmath}
\usepackage{amssymb}
\usepackage{amsthm}
\usepackage{amsfonts}
\usepackage{listings}
\usepackage{enumerate}
\usepackage{latexsym}
\usepackage{psfrag}
\usepackage{bm}
\usepackage{bbm}
\usepackage{hyperref}

\newcommand{\beq}{\begin{equation}}
\newcommand{\eeq}{\end{equation}}
\newcommand{\bea}{\begin{eqnarray}}
\newcommand{\eea}{\end{eqnarray}}

\newcommand{\eq}[1]{Eq.~(\ref{eq:#1})}
\newcommand{\eqs}[2]{Eqs.~(\ref{eq:#1}) and~(\ref{eq:#2})}
\newcommand{\equ}[1]{Equation~(\ref{eq:#1})}
\newcommand{\equs}[2]{Equations~(\ref{eq:#1}) and (\ref{eq:#2})}

\def\nn{\nonumber\\}

\def\me#1#2#3{\langle#1\vert#2\vert#3\rangle}

\def\w{\omega}
\def\q{{\bf q}}
\def\M{{\bf M}}
\def\P{{\bf P}}
\def\J{{\bm j}}
\def\A{{\bf A}}

\def\B{{\bf B}}
\def\E{{\bf E}}
\def\bb{^{\rm{\bf B}}}
\def\ee{^{\rm{\bf E}}}
\def\r{{\bf r}}
\def\k{{\bf k}}
\def\kf{\k_F}
\def\sta{^{\text{stat}}}
\def\cme{^{\text{CME}}}
\def\gme{^{\text{GME}}}
\def\re{{\text{Re}}}
\def\im{{\text{Im}}}

\def\a{^{A}}
\def\dkkk{[d\k]}

\def\e{\epsilon}

\def\ef{\epsilon_F}
\def\eavg{\overline{\epsilon}}

\def\d{{\bf d}}

\def\ij{(i\leftrightarrow j)}
\def\vv{{\bf v}}
                   
\def\vf{\hat{\vv}_F}
\def\vfmag{v_F}

\def\curv{{\boldsymbol\Omega}}

\def\bpartial{{\boldsymbol\partial}}
\def\R{_R}
\def\L{_L}
\def\muR{\mu\R}
\def\muL{\mu\L}
\def\eR{\epsilon\R}
\def\eL{\epsilon\L}

\def\balpha{{\boldsymbol\alpha}}
\def\egap{\epsilon_{\text{gap}}}
\def\ME{^{\text{me}}}
\def\EM{^{\text{em}}}
\def\ident{\mathbbm{1}}

\begin{document}

%===========================%
% TITLE PAGE                %
%===========================%
\title{ Gyrotropic Magnetic Effect and the Magnetic Moment on the
  Fermi Surface}
\author{Shudan Zhong} 
\affiliation{Department of Physics, University of California,
  Berkeley, California 94720, USA}
\author{Joel E.  Moore} 
\affiliation{Department of Physics, University of California,
  Berkeley, California 94720, USA}
\affiliation{Materials Sciences Division, Lawrence Berkeley National
  Laboratory, Berkeley, California 94720, USA}
\author{Ivo Souza} 
\affiliation{Centro de F\'{\i}sica de Materiales, Universidad del
  Pa\'{\i}s Vasco, 20018 San Sebast\'ian, Spain}
\affiliation{Ikerbasque Foundation, 48013 Bilbao, Spain}

\date{\today}
\begin{abstract}

  The current density $\J\bb$ induced in a clean metal by a
  slowly-varying magnetic field $\B$ is formulated as the
  low-frequency limit of natural optical activity, or natural
  gyrotropy.  Working with a multiband Pauli Hamiltonian, we obtain
  from the Kubo formula a simple expression for
  $\alpha\gme_{ij}=j\bb_i/B_j$ in terms of the intrinsic magnetic
  moment (orbital plus spin) of the Bloch electrons on the Fermi
  surface. An alternate semiclassical derivation provides an intuitive
  picture of the effect, and takes into account the influence of
  scattering processes in dirty metals.  This ``gyrotropic magnetic
  effect'' is fundamentally different from the chiral magnetic effect
  driven by the chiral anomaly and governed by the Berry curvature on
  the Fermi surface, and the two effects are compared for a minimal
  model of a Weyl semimetal.  Like the Berry curvature, the intrinsic
  magnetic moment should be regarded as a basic ingredient in the
  Fermi-liquid description of transport in broken-symmetry metals.

\end{abstract}
% \pacs{78.20.Ek,75.47.-m,71.18.+y}
\maketitle

%===========================%
% MAIN TEXT                 %
%===========================%

%---------------------
{\it Introduction.---}  
%---------------------
%
When a solid is placed in a static magnetic field the nature of the
electronic ground state can change, leading to striking transport
effects. A prime example is the integer quantum Hall effect in a
quasi-two-dimensional metal in a strong perpendicular
field~\cite{thouless-prl82}.  Novel magnetotransport effects have also
been predicted to occur in 3D topological (Weyl) metals, such as an
anomalous longitudinal
magnetoresistence~\cite{nielsen-pl83,son-prb13}, and the chiral
magnetic effect (CME), where an electric pulse $\E\parallel \B$
induces a transient current $\J\parallel \B$~\cite{son-prl12}; both
are related to the chiral anomaly that was originally discussed for
Weyl fermions in particle physics~\cite{adler-pr69,bell-nc69}.  In all
these phenomena the role of the static $\B$~field is to modify the
equilibrium state, but an $\E$~field is still required to put the
electrons out of equilibrium and drive the current (since
$\E=-\dot{\A}$, the vector potential is time dependent even for a
static $\E$~field).

Recently, the intriguing proposal was made that a pure $\B$~field
could drive a dissipationless current in certain Weyl semimetals where
isolated band touchings [the ``Weyl points'' (WPs)] of opposite
chirality are at different energies~\cite{zyuzin-prb12}.  The
existence of such an effect was later questioned~\cite{vazifeh-prl13},
and the initial interpretation as an {\it equilibrium} current was
discounted.  (Indeed, that would a violate a ``no-go theorem''
attributed to Bloch that forbids macroscopic current in a bulk system
in equilibrium~\cite{yamamoto-prd15}.) Subsequent theoretical work
suggests that the proposed effect can still occur in {\it transport},
as the current response to a $\B$~field oscillating at low
frequencies~\cite{chen-prb13,goswami-arxiv13,chang-prb15,chang-prb15b}.

At present the effect is still widely regarded as being related to the
chiral anomaly~\cite{chen-prb13} (or, more generally, to the Berry
curvature of the Bloch
bands~\cite{goswami-arxiv13,chang-prb15,chang-prb15b,goswami-prb15}),
and is broadly characterized as a type of CME.  We show in this Letter
that the experimental implications and microscopic origin of this
effect are both very different from the CME (as defined in
Ref.~\cite{son-prl12}, consistent with the particle-physics
literature~\cite{kharzeev-ppnp14}). Experimentally, the effect is
realized as the low-frequency limit of natural gyrotropy~\footnote{The
  term {\it natural gyrotropy} refers to the time-reversal-even part
  of the optical response of a medium at linear order in the wave
  vector of light~\cite{landau-EM,agranovich-book84}. The reactive
  part gives rise to natural optical rotation, and the dissipative
  part to natural circular dichroism. Furthermore, polar crystals
  display natural gyrotropy effects unrelated to optical
  rotation~\cite{agranovich-book84}. Gyrotropic effects that are
  time-reversal-odd and zeroth order in the wave vector of light
  (e.g., Faraday rotation and magnetic circular
  dichroism~\cite{landau-EM}) are not considered in this work.} in
clean metals (see also Ref.~\onlinecite{goswami-prb15}), and we will
call it the ``gyrotropic magnetic effect'' (GME). Both $\E$ and~$\B$
optical fields drive the gyrotropic current, but at frequencies well
below the threshold for interband absorption ($\hbar\w\ll\egap$) their
separate contributions can be identified. In nonpolar metals, the
induced gyrotropic current can be inferred from optical rotation
measurements.  The GME is predicted to occur not only in certain Weyl
semimetals, but in any optically active metal; it is necessary that
the structure lacks an inversion center, and it is sufficient that the
structure is either chiral~\cite{landau-EM,flack-hca03,newnham-book05}
or polar~\cite{agranovich-book84}.

Existing expressions for the natural gyrotropy current in metals
involve the Berry curvature of all the occupied states (and velocities
of empty
bands)~\cite{goswami-arxiv13,chang-prb15,chang-prb15b,goswami-prb15},
at odds with the notion that transport currents are carried by states
near the Fermi level $\ef$.  Integrals over all occupied states
involving the Berry curvature also appear in calculations of a part of
the low-frequency optical
activity~\cite{orenstein-prb12,hosur-prb15,zhong-prl15}, and of the
anomalous Hall effect (AHE); in the case of the AHE, a Fermi surface
(FS) reformulation exists~\cite{haldane-prl04}.  We find that the GME
is not governed by the chiral anomaly or the Berry curvature, but by
the intrinsic magnetic moment of the Bloch states on the FS.  Our
analysis also takes into account the finite relaxation time $\tau$ in
real materials, which is shown to weaken the effect at the lowest
frequencies. The magnitude of the GME is estimated for the predicted
chiral Weyl semimetal SrSi$_2$~\cite{huang-arxiv15}.

\begin{figure*}
\begin{center}
\includegraphics[width=0.55\columnwidth]{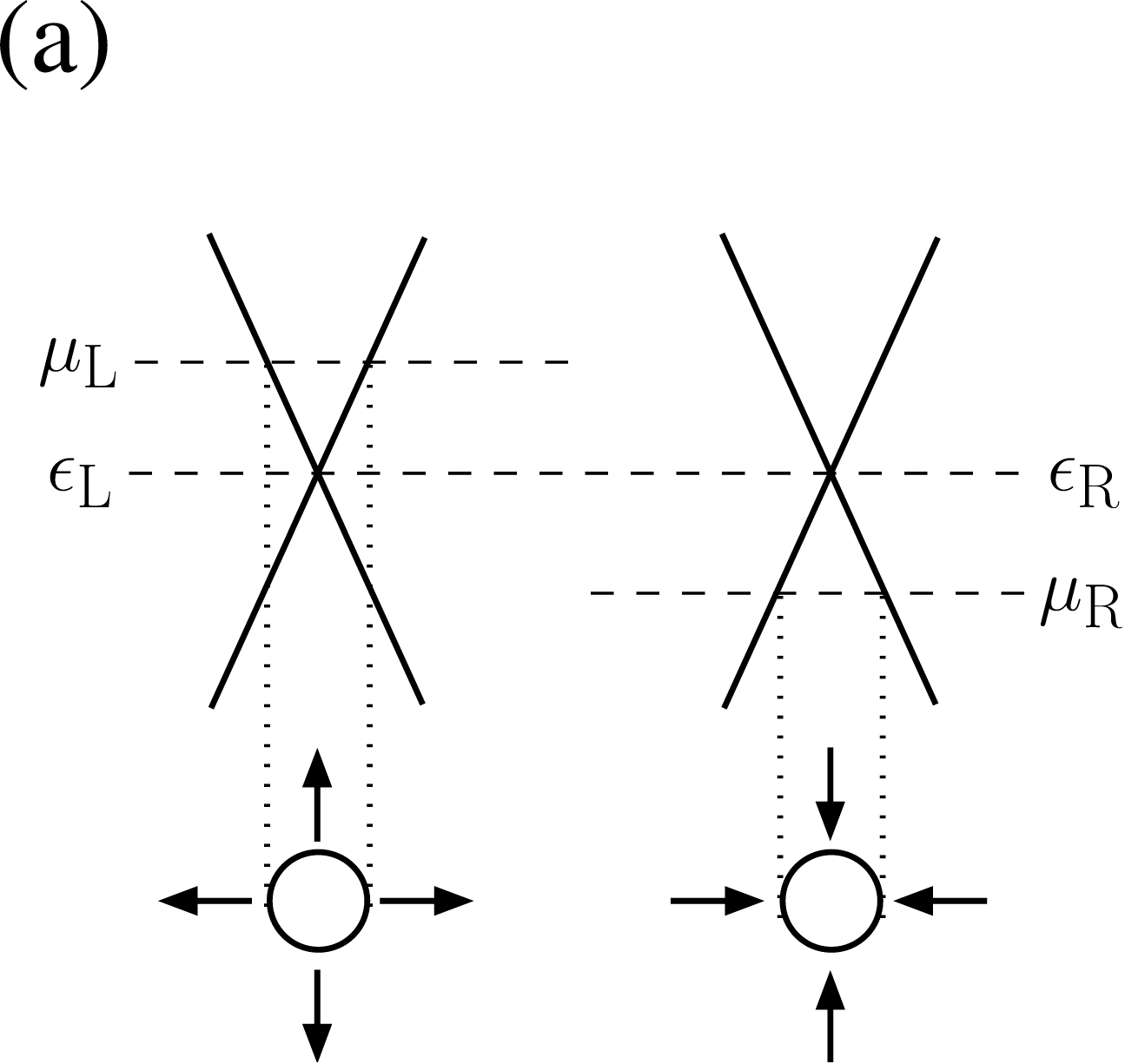}
\hspace{1cm}
\includegraphics[width=0.55\columnwidth]{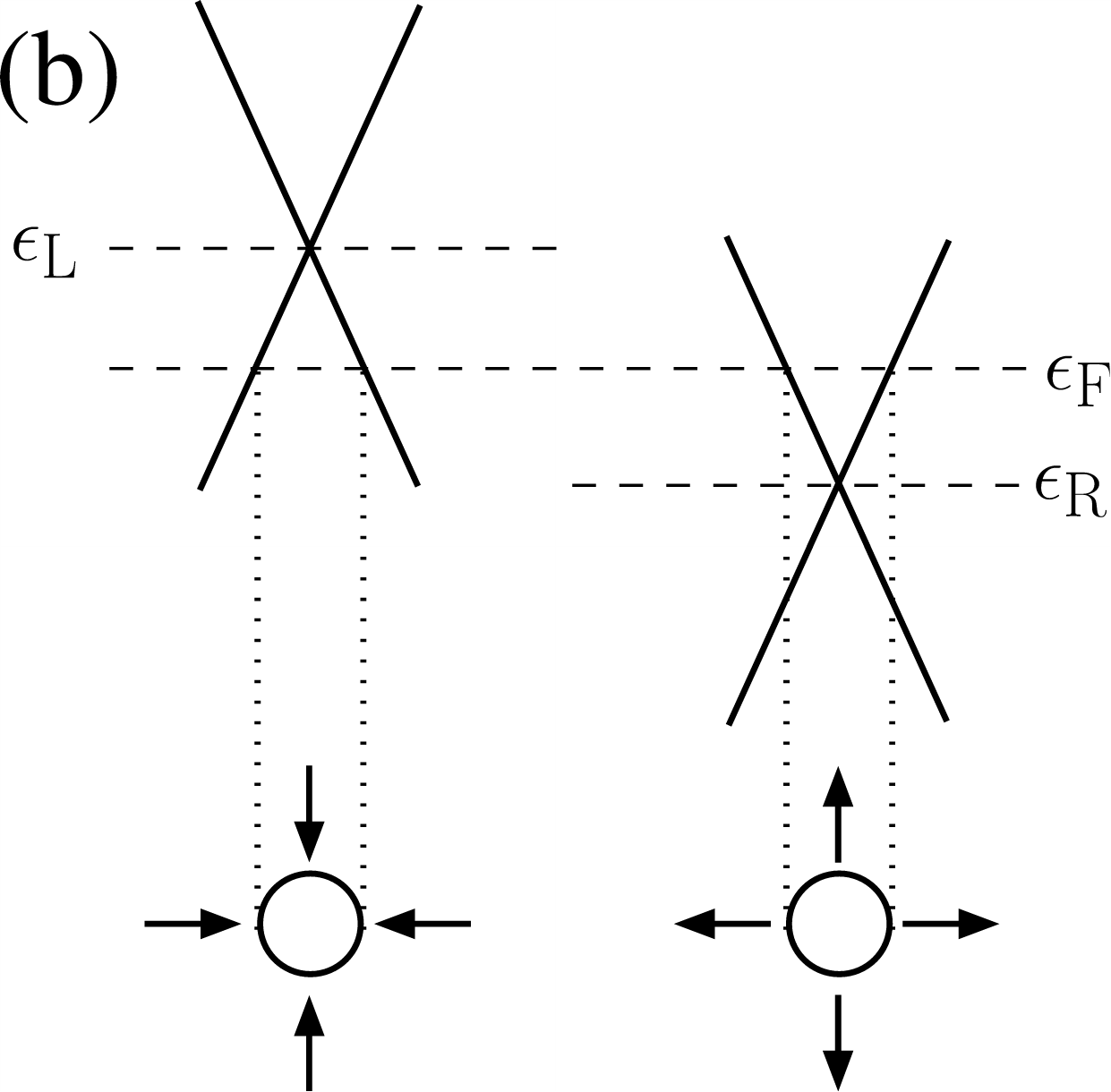}
\end{center}
\caption{ (a) Chiral magnetic effect in a $T$-broken Weyl semimetal in
  a static $\B$~field. The left- and right-handed Weyl nodes are at
  the same energy $\e\L=\e\R$, but the enclosing Fermi pockets are not
  in chemical equilibrium ($\muL\not=\muR$) due to the application of
  an $\E\parallel\B$ pulse, and this drives the current
  [\eq{J-cme-2}].  (b) Gyrotropic magnetic effect. $P$ symmetry is now
  broken along with~$T$, leading to $\e\L\not=\e\R$.  The Fermi
  pockets are in chemical equilibrium, $\muL=\muR=\ef$, and an
  oscillating $\B$~field drives the current [\eq{J-gme-2}].  The
  bottom of each panel shows the Fermi pockets, and the arrows
  represent the Fermi velocities.}
\label{fig}
\end{figure*}

%-----------------------
{\it CME versus GME.---}  
%-----------------------
%
Both effects can be discussed by positing a linear relation between
$\J$ and $\B$:
\beq
\label{eq:J-B}
j_i=\alpha_{ij}B_j.
\eeq
Suppose we use linear response to evaluate $\balpha$ for a clean
metal, describing the $\B$~field in terms of a vector potential that
depends on both $\q$ and $\w$. The result will depend on the order in
which the $\q\rightarrow 0$ and $\w\rightarrow 0$ limits are
taken~\cite{chen-prb13,goswami-arxiv13,chang-prb15}, much as the
compressibility and conductivity are different limits of electrical
response.  The CME tensor $\balpha\cme$ can be obtained from \eq{J-B}
in the equilibrium or {\it static} limit of the magnetic field
(setting $\w=0$ before sending $\q\rightarrow 0$), with an additional
step needed to describe the $\E$-field pulse.  The GME tensor
$\balpha\gme$ is extracted directly from \eq{J-B} in the transport or
{\it uniform} limit (sending $\q\rightarrow 0$ before $\w\rightarrow
0$) that describes conductivities in experiment. (Here,
``$\w\rightarrow 0$'' means $\hbar\w\ll\egap$, but note that
$\w\tau\gg 1$ because the clean limit $\tau\rightarrow\infty$ is
assumed; effects caused by finite relaxation times in dirty samples
will be discussed later.)  Only $\balpha\gme$ is a %an intrinsic
material property, since the details of the $\E$-field pulse producing
nonequilibrium are missing from $\balpha\cme$.  Below we derive
microscopic expressions for both. %each of them in turn.

%-------------------------------
{\it Chiral magnetic effect.---}  
%-------------------------------
%
The tensor $\balpha$ calculated in the static limit is isotropic,
$\alpha_{ij}=\alpha\sta\delta_{ij}$, with
\beq
\label{eq:alpha-sta}
\alpha\sta=-\frac{e^2}{\hbar}\sum_n\int\dkkk\, f^0_{\k n}
\left(\vv_{\k n}\cdot\curv_{\k n}\right)=0\,,
\eeq 
where $\dkkk=d^3k/(2\pi)^3$, the integral is over the Brillouin zone,
$f^0_{\k n}=f(\e_{\k n})$ is the equilibrium occupation factor,
$\vv_{\k n}= \bpartial_{\hbar \k} \e_{\k n}$ is the band velocity,
$\curv_{\k n}=-\im\me{\bpartial_\k u_{\k n}}{\times}{\bpartial_\k
  u_{\k n}}$ is the Berry curvature, and $-e$ is the electron
charge. \equ{alpha-sta} was derived in Ref.~\cite{zhou-cpl13} using
the semiclassical formalism~\cite{xiao-rmp10}, and we obtain the same
result from linear response~\footnote{See Supplemental Material at
  \url{http://cmt.berkeley.edu/suppl/zhong-arxiv15-suppl.pdf}, which
  includes Refs.~[29-43], for (i) the derivation of
  \eqs{alpha-sta}{Pi-A-ijl} from linear response, (ii) the derivation
  from \eq{gme-pre} of the non-FS formula for $\overline{\alpha}\gme$
  given in Refs.~\cite{goswami-arxiv13,chang-prb15}, (iii) the
  derivation of \eq{rotatory-power} from \eq{J-gme-2} combined with
  phenomenological relations, (iv) an analysis of the gyrotropic
  response in polar metals, and of the gyrotropic response induced in
  Weyl semimetals by the chiral anomaly, (v) the derivation of a
  reciprocity relation for the natural gyrotropy of a metal with a
  smooth interface, and (vi) the identification of a traceless
  Berry-curvature piece in the full GME response tensor of
  \eq{gme-pre}.}.  The fact that $\alpha\sta$ vanishes (see below) is
in accord with Bloch's theorem~\cite{yamamoto-prd15}.

To turn the above ``quasiresponse'' into $\alpha\cme$, let us recast
\eq{alpha-sta} as a FS integral.  Integrating by parts produces two
terms.  The one containing $\bpartial_\k\cdot\curv_{\k n}$ picks up
monopole contributions from the occupied WPs, and vanishes because
each WP appears twice with opposite signs~\cite{gosalbez-prb15}.  In
the remaining term we write $\bpartial_\k f^0=-\vf\delta^3(\k-\kf)$,
with $\vf$ the FS normal at $\kf$, and introduce the Chern number
$C_{na}=(1/2\pi)\int_{S_{na}}dS\left(\vf\cdot\curv_{\k n}\right)$ of
the $a$th Fermi sheet $S_{na}$ in band
$n$~\cite{haldane-prl04,gosalbez-prb15}. After assigning different
chemical potentials to different sheets to account for the effect of
the $\E$-field pulse, \eq{alpha-sta} becomes
$\alpha\cme=-(e^2/h^2)\sum_{n,a}\mu_{na}C_{na}$, leading to the
current density $\J=\alpha\cme\B$~\cite{son-prl12,yamamoto-prd15}.  In
equilibrium $\mu_{na}=\ef$, and using $\sum_{n,a}C_{na}=0$ we find
$\J=0$, as per \eq{alpha-sta}.

For a Weyl semimetal with two Fermi pockets with $C=+1$ and $C=-1$
placed at slightly different chemical potentials $\muL$ and
$\muR$~\footnote{With our sign convention for the Berry curvature, a
  right-handed WP acts as a source in the lower band and as a sink in
  the upper band~\cite{xiao-rmp10}.  An enclosing pocket, either
  electronlike or holelike, has Chern number $C=-1$.}
[Fig.~\ref{fig}(a)], a current develops:
\beq
\label{eq:J-cme-2}
\J=(e^2/h^2)\B(\muR-\muL)\,.
\eeq

%-----------------------------------
{\it Gyrotropic magnetic effect.---}  
%-----------------------------------
%
Symmetry considerations already suggest a link between the GME and
natural gyrotropy.  Both $\J$ and $\B$ are odd under time
reversal~$T$, and $\J$ is odd under spatial inversion~$P$, while $\B$
is $P$~even, and so according to \eq{J-B} the GME is $T$~even and
$P$~odd, the same as natural gyrotropy~\cite{Note1}.

To make the connection precise, consider the current density induced
by a monochromatic electromagnetic field
$\A(t,\r)=\A(\w,\q)e^{i(\q\cdot\r-\w t)}$ at first order in $\q$:
\beq
j_i(\w,\q)=\Pi_{ijl}(\w)A_j(\w,\q)q_l\,.
\eeq
The $T$-even part $\Pi\a_{ijl}$ of the response tensor is
antisymmetric ($A$) under $i\leftrightarrow j$. It has nine
independent components, and can be repackaged as a rank-2 tensor
using~\cite{hornreich-pr68,malashevich-prb10}
\begin{subequations}
\label{eq:Pi2alpha}
\bea
\label{eq:Pi-alpha}
\Pi\a_{ijl}&=&i\varepsilon_{ilp}\alpha\gme_{jp}
-i\varepsilon_{jlp}\alpha\gme_{ip}\,\\
\label{eq:alpha-Pi}
\alpha\gme_{ij}&=&\frac{1}{4i}\varepsilon_{jlp}
\left(\Pi\a_{lpi}-2\Pi\a_{ilp}\right)\,.
\eea
\end{subequations}
At nonabsorbing frequencies $\balpha\gme(\w)$ is real and
${\boldsymbol\Pi}\a(\w)$ is purely imaginary, but otherwise both are
complex.

From now on we assume $\hbar\w\ll\egap$, so that only intraband
absorption can occur.  In this regime $\balpha\gme$ satisfies
\begin{subequations}
\label{eq:gme-direct-inverse}
\bea
\label{eq:gme-direct}
j\bb_i&=&-i\w P\bb_i=\alpha\gme_{ij}B_j\\
\label{eq:gme-inverse}
M\ee_i&=&-(i/\w)\alpha\gme_{ji}E_j\,,
\eea
\end{subequations}
where $\E=i\w\A$ and $\B=i\q\times\A$, and $\P\bb$ and $\M\ee$ are
oscillating moments induced by $\B$ and $\E$ respectively.  The
natural gyrotropy current 
is $\J\bb+i\q\times\M\ee$.
% $j^g_i=\Pi\a_{ijl}A_jq_l=j\bb_i+j\ee_i$ has
% contributions from both $\B$ and $\E$, with $\J\bb$ given by
% \eq{gme-direct} and $\J\ee=i\q\times\M\ee$.
%
In the long-wavelength limit \eq{gme-direct} describes a transport
current induced by a time-varying $\B$ in an optically active metal
(the {\it direct} GME), and \eq{gme-inverse} describes a macroscopic
magnetization induced by~$\E$; this {\it inverse} GME has been
previously discussed for polar~\cite{edelstein-prb11} and
chiral~\cite{yoda-sr15} metals.

To derive \eq{gme-direct-inverse}, consider a finite sample of size
$L$.  Using Eq.~(20) of Ref.~\cite{malashevich-prb10} for
$\sigma\a_{ijl}=(1/i\w)\Pi\a_{ijl}$ we find~\footnote{To recover the
  bulk result from \eq{alpha-finite}, the $L\rightarrow\infty$ limit
  should be taken faster than the $\w\rightarrow 0$ limit, consistent
  with the order of limits discussed earlier for transport.}
\beq
\label{eq:alpha-finite}
\alpha\gme_{ij}=
% -i\w\frac{1}{2}
(\w/2i)
\left(\chi\EM_{ij}-\chi\ME_{ji}\right)+
\text{(E.Q. terms)\,.}
\eeq 
``E.Q.'' denotes electric quadrupole terms that keep $\balpha\gme$
origin independent at higher
frequencies~\cite{buckingham-jcs71,malashevich-prb10}, but do not
contribute to $\J\bb$ or $\M\ee$ when $\hbar\w\ll \egap$, as they are
higher order in $\w$ than the first term. The low-frequency gyrotropic
response is controlled by the magnetoelectric susceptibilities
$\chi\EM_{ij}=\partial P_i/\partial B_j$ and $\chi\ME_{ij}=\partial
M_i/\partial E_j$.  The dynamic polarization $P\bb_i$ can be
decomposed into $T$-even and $T$-odd parts
$(1/2)(\chi\EM_{ij}-\chi\ME_{ji})B_j$ and
$(1/2)(\chi\EM_{ij}+\chi\ME_{ji})B_j$~\footnote{This decomposition is
  obtained by invoking the Onsager relation
  $\left.\chi\EM_{ij}(\w)\right|_{-\B_{\rm
      ext}}=-\left.\chi\ME_{ji}(\w)\right|_{\B_{\rm
      ext}}$~\cite{melrose-book91}.}, and \eq{gme-direct} corresponds
to the former. Similarly, \eq{gme-inverse} gives the $T$-even part of
the magnetization induced by $\E$.  (The $T$-odd part of the
magnetoelectric susceptibilities describes the linear magnetoelectric
effect in insulators such as Cr$_2$O$_3$.)

In brief, the GME is the low-frequency limit of natural gyrotropy in
$P$-broken metals, in much the same way that the AHE is the transport
limit of Faraday rotation in $T$-broken metals.  While the intrinsic
AHE is governed by the geometric Berry
curvature~\cite{xiao-rmp10,haldane-prl04} and becomes quantized by
topology in Chern insulators, the GME is controlled by a nongeometric
quantity, the intrinsic magnetic moment of the Bloch states on the
FS~\footnote{Here, the term {\it geometric} refers to the intrinsic
  geometry of the Bloch-state fiber bundle. The orbital moment of
  Bloch electrons can be considered geometric in a different sense: it
  is the imaginary part of a complex tensor whose real part gives the
  inverse effective mass tensor, i.e., the curvature of band
  dispersions~\cite{gao-prb15}.}.

To establish this result let us return to periodic crystals and derive
a bulk formula for $\balpha\gme$ at $\hbar\w\ll\egap$. From the Kubo
linear response in the uniform limit, we obtain~\cite{Note2}
\begin{eqnarray}
\label{eq:Pi-A-ijl}
\Pi\a_{ijl}&=&
\frac{e^2\w\tau}{1-i\w\tau}
\sum_n\int\dkkk\,
\frac{\partial f}{\partial\e_{\k n}}
\Big[
-\frac{g_s}{2m_e}\varepsilon_{ipl}v_{\k n,j}S_{\k n,p}  \nn
  &+&\frac{v_{\k n,i}}{\hbar}\im\me{\partial_j u_{\k n}}{H_\k - \e_{\k n}}{\partial_l u_{\k n}}
  -\ij 
\Big].
\end{eqnarray}
[The calculation was carried out for a clean metal where formally
$\tau=1/\eta$ and $\eta\rightarrow 0^+$~\cite{allen-cccms06}.
Alternately one could retain a finite $\tau$ to give a
phenomenological relaxation time in dirty metals, and indeed the
semiclassical relaxation-time calculation to be presented shortly
gives the same Drude-like dependence on $\omega \tau$ as
\eq{Pi-A-ijl}.] ${\bf S}_{\k n}$ is the expectation value of the spin
${\bf S}=(\hbar/2){\boldsymbol\sigma}$ of a Bloch state, $g_s\simeq2$
is the spin $g$~factor of the electron, and $m_e$ is the electron
mass. Inserting \eq{Pi-A-ijl} into \eq{alpha-Pi} gives
\beq
\label{eq:gme-pre}
\alpha\gme_{ij}=
\frac{i\w\tau e}{1-i\w\tau}
\sum_n\int\dkkk\,(\partial f/\partial\e_{\k n})
v_{\k n,i}m_{\k n,j}\,,
\eeq
where ${\bf m}_{\k n}=-(eg_s/2m_e){\bf S}_{\k n}+{\bf m}^{\rm orb}_{\k
  n}$ is the magnetic moment of a Bloch electron, whose orbital part
is~\cite{xiao-rmp10}
\beq
\label{eq:m-orb} {\bf m}^{\rm orb}_{\k n}=\frac{e}{2\hbar}
\im\me{\bpartial_\k u_{\k n}}{\times(H_\k-\e_{\k n})}{\bpartial_\k
  u_{\k n}}\,. 
\eeq
At zero temperature, we can replace $\partial f/\partial\e_{\k n}$ in
\eq{gme-pre} with $-\delta^3(\k-\kf)/\hbar|\vv_{\k n}|$ to obtain the
FS formula
\beq
\label{eq:gme}
\alpha\gme_{ij}=
\frac{i\w\tau}{i\w\tau-1}\frac{e}{(2\pi)^2h}
\sum_{n,a}\int_{S_{na}} 
dS\,\hat{v}_{F,i}m_{\k n,j}\,.
\eeq
A nonzero ${\bf m}_{\k n}$ requires broken $PT$ symmetry, but the GME
can only occur if $P$ is broken: with $P$ symmetry present ${\bf
  m}_{-\k,n}={\bf m}_{\k n}$ and $\vf(-\kf)=-\vf(\kf)$, leading to
$\balpha\gme=0$.  Without spin-orbit coupling, only the orbital moment
contributes.

\equs{gme-direct-inverse}{gme} are our main results. The GME is fully
controlled by the bulk FS and vanishes trivially for insulators,
contrary to the AHE where the FS formulation misses possible quantized
contributions~\cite{haldane-prl04}.

According to \eq{gme}, the reactive response $\re\balpha\gme$ is
suppressed by scattering when $\w\ll 1/\tau$. It increases with~$\w$,
and levels off for $\w\gg 1/\tau$ (satisfying this condition without
violating $\hbar\w\ll\egap$ requires sufficiently clean samples). The
opposite is true for the dissipative response $\im\balpha\gme$, which
drops to zero at $\w\gg 1/\tau$ and becomes strongest at $\w\ll
1/\tau$.  In this lowest-frequency limit $\J\bb\rightarrow 0$, and
\eqs{gme-inverse}{gme-pre} for the induced magnetization reduce to the
expression in Ref.~\cite{yoda-sr15}. Thus, in the dc limit only a
dissipative inverse GME occurs in dirty metals.

%------------------------------
{\it Semiclassical picture.---} 
%------------------------------
%
Our discussion of the GME assumed from the outset
$\hbar\w\ll\egap$. Since this is the regime where the semiclassical
description of transport in metals holds~\cite{ashcroft-book76}, it is
instructive to rederive \eqs{gme-direct-inverse}{gme} by solving the
Boltzmann equation. This provides an intuitive picture of the GME and
its modification by scattering processes.  The key ingredient beyond
previous semiclassical
approaches~\cite{orenstein-prb12,hosur-prb15,zhong-prl15} is the
correction to the band energy and the band velocity (as opposed to the
Berry-curvature anomalous velocity) in the presence of a magnetic
field~\cite{chang-prb15,xiao-rmp10}: $\tilde{\vv}_{\k n} =
\bpartial_{\hbar\k} \tilde{\epsilon}_{\k n}$, where
$\tilde{\epsilon}_{\k n}=\e_{\k n}-{\bf m}_{\k n} \cdot \B$.

In a static $\B$~field, the conduction electrons reach a new
equilibrium state with $f^0_{\k n}(\B)=f(\tilde{\epsilon}_{\k n})$ as
the distribution function~\cite{chang-prb15}, and the current vanishes
according to \eq{alpha-sta}.  Under oscillating fields $\E,\,\B\propto
e^{i(\q\cdot\r-\w t)}$ the electrons are in an excited state with a
distribution function $g_{\k n}(t,\r)$ which we find by solving the
Boltzmann equation in the relaxation-time approximation,
\beq
\partial_t g_{\k n}
+\dot{\r}\frac{\partial g_{\k n}}{\partial\r} 
+\dot{\k}\frac{\partial g_{\k n}}{\partial\k} 
= -\left[g_{\k n} - f_{\k n}^0(\B)
\right]/\tau\,,
\eeq 
where $\tau$ is the relaxation time to return to the instantaneous
equilibrium state described by $f^0_{\k n}\left(\B(t,\r)\right)$ (for
a slow spatial variation of $\B$). Using the semiclassical
equations~\cite{xiao-rmp10}, the distribution function to linear order
in $\E$ and $\B$ is $g_{\k n}(t,\r) = f^0_{\k n}\left(\B(t,\r)\right)
+ f^1_{\k n}(t,\r)$, with
\beq
\label{eq:distribution}
f^1_{\k n} = 
\frac{\partial f/\partial \e_{\k n}}{1-\frac{\q}{\omega} 
\cdot \vv_{\k n} + \frac{i}{\omega \tau}}
\Big[ {\bf m}_{\k n} \cdot \B + (ie/\w) \E \cdot \vv_{\k n} \Big] 
\,,
\eeq
which at $\w\tau\gg 1$ reduces to the result in
Ref.~\onlinecite{chang-prb15}.

As the current associated with $f^0_{\k n}(\B)$ vanishes, the current
induced by an oscillating $\B$~field is obtained by multiplying the
first term in \eq{distribution} with the unperturbed band velocity.
The result in the long-wavelength limit is
\beq 
\J\bb = \frac{i\omega\tau e}{1 - i\omega \tau} \sum_n
\int \dkkk\, 
(\partial f/\partial \e_{\k n})\,
\vv_{\k n}\,({\bf m}_{\k n} \cdot \B)\,, 
\eeq
in agreement with \eqs{gme-direct}{gme-pre}. Conversely, inserting the
second term of \eq{distribution} in the bulk expression for
$\M=\M^{\rm spin}+\M^{\rm orb}$~\cite{xiao-rmp10} leads to
\eqs{gme-inverse}{gme-pre} for the magnetization induced by an
oscillating $\E$~field.

%-------------------------------
{\it GME in two-band models.---}  
%-------------------------------
%
Consider a situation where only two bands are close to $\ef$, and
couplings to more distant bands can be neglected when evaluating the
orbital moment on the FS (for simplicity, we focus here on the orbital
contribution). The Hamiltonian written in the basis of the identity
matrix and the three Pauli matrices is
$H_\k=\eavg_\k\ident+\d_\k\cdot{\boldsymbol\sigma}$, with eigenvalues
$\e_{\k t}=\eavg_\k+td_\k$, where $t=\pm 1$ and
$d_\k=|\d_\k|$. \equ{m-orb} becomes
\beq
\label{eq:m}
m^{\rm orb}_{\k t,i}
=-\frac{e}{\hbar}\varepsilon_{ijl}\frac{1}{2d_\k^2}
\d_\k\cdot\left(\partial_j\d_\k\times\partial_l \d_\k\right)\,.
\eeq

For orientation we study a minimal model for a Weyl semimetal where
the FS consists of two pockets surrounding isotropic WPs of opposite
chirality. We allow the WPs to be at different energies (this requires
breaking both $P$ and $T$), but $\ef$ is assumed close to both
[Fig.~\ref{fig}(b)]. Near each WP the Hamiltonian is
$H_{\k\nu}=\epsilon_\nu\ident+\chi_\nu\hbar
\vfmag\k\cdot{\boldsymbol\sigma}$, where $\nu$ labels the WP,
$\epsilon_\nu$ and $\chi_\nu=\pm 1$ are its energy and chirality
(positive means right-handed), $\k$ is measured from the WP, and
$\vfmag$ is the Fermi velocity. From \eq{m}, ${\bf m}^{\rm
  orb}_{\k\nu}=-\chi_\nu(e\vfmag/2k)\hat{\k}$ for $t=\pm 1$, and only
the trace piece $\overline{\alpha}\gme\delta_{ij}$ survives is
\eq{gme}; in the clean limit each pocket contributes
\beq
\overline{\alpha}\gme_\nu
=\mp\frac{1}{3}\frac{e^2}{h^2}\chi_\nu\hbar\vfmag k_{\rm F}
=\frac{1}{3}\frac{e^2}{h^2}\chi_\nu(\epsilon_\nu-\ef)\,,
\eeq
where the minus (plus) sign in the middle expression corresponds to
$\epsilon_\nu<\ef$ ($\epsilon_\nu>\ef$). Summing over $\nu$ and using
$\sum_\nu\chi_\nu=0$~\cite{nielsen-np81} gives $\overline{\alpha}\gme
=(e^2/3h^2)\sum_\nu\chi_\nu\epsilon_\nu$. For a minimal model
$\nu=L,R$, and the GME current is
\beq
\label{eq:J-gme-2}
\J\bb=(e^2/3h^2)(\epsilon_R-\epsilon_L)\B\,.  
\eeq

\equ{J-gme-2} looks deceptively similar to \eq{J-cme-2} for the CME
current.  The prefactor is different, but the key difference is in the
meaning of the various quantities, and in their respective roles.  To
stress this point, in both equations we have placed the ``force'' that
drives the current at the end, after the equilibrium parameter that
enables the effect.  The GME current is driven by the oscillating
$\B$~field, while $\eL$ and $\eR$ are band structure parameters, with
$\eR-\eL$ reflecting the degree of structural symmetry breaking that
allows the effect to occur.  \equ{J-cme-2} is ``universal'' because of
the topological nature of the FS integral involved, while \eq{J-gme-2}
is for spherical pockets surrounding isotropic Weyl nodes. For generic
two-band models the traceless part of $\balpha\gme$ is generally
nonzero~\footnote{For any number of bands, the traceless part of
  $\balpha\gme$ includes~\cite{Note2} the Berry-curvature piece found
  previously~\cite{zhong-prl15}, and the full tensor
  satisfies~\cite{Note2} the microscopic constraint from time-reversal
  symmetry previously shown for that traceless
  piece~\cite{zhong-prl15}.}, and the non-FS expression of
Refs.~\cite{goswami-arxiv13,chang-prb15} for the orbital contribution
to the trace can be recovered from \eq{gme-pre}~\cite{Note2}.
  
We emphasize that breaking $T$ is not required for the GME. If~$T$ is
present (and $P$ broken), the minimum number of WPs is four, not
two~\cite{young-prl12}.  In the class of $T$-symmetric Weyl materials
so far discovered, $T$ relates WPs of the same chirality and energy.
Mirror symmetries connect WPs of opposite chirality so that 
% $\J\bb=0$,
${\bm j}^{\rm{\bf B}}\cdot{\rm{\bf B}}=0$, as expected since these
symmetries tend to exclude optical
rotation~\cite{flack-hca03,newnham-book05}. Fortuitously, the
predicted Weyl material SrSi$_2$ has misaligned WPs of opposite
chirality due to broken mirror symmetry~\cite{huang-arxiv15}.  Its
rotatory power~$\rho$ can be estimated from the energy splitting
between WPs. Neglecting anisotropy effects and spin contributions that
were not included in \eq{J-gme-2}, each WP pair
contributes~\cite{Note2}
\beq
\label{eq:rotatory-power}
\rho=(2\alpha/3hc)\left(\epsilon_L-\epsilon_R\right)\,,
\eeq
with $\alpha$ the fine-structure constant and $c$ the speed of
light. The calculated splitting $|\epsilon_L-\epsilon_R|\sim
0.1$~eV~\cite{huang-arxiv15} gives $|\rho|\sim 0.4$~rad/mm per node
pair, about the same as $|\rho|=0.328$~rad/mm for quartz at
$\lambda=0.63$~$\mu$m~\cite{newnham-book05}.  This should be
measurable in a frequency range from the infrared (above which the
semiclassical assumptions break down) down to $1/\tau$, which depends
on crystal quality. When $\epsilon_L=\epsilon_R$ the rotatory power
vanishes in equilibrium, but a nonequilibrium gyrotropic effect can
still occur due to the chiral anomaly~\cite{hosur-prb15,Note2}.  In
polar metals, the tensor $\balpha\gme$ acquires am antisymmetric part
(equivalent to a polar vector ${\boldsymbol\delta}$) that does not
contribute to optical rotation, but which leads to a {\it transverse}
GME of the form
$\M\ee\propto\E\times{\boldsymbol\delta}$~\cite{Note2}.

In summary, we have elucidated the physical origin of currents induced
by low-frequency magnetic fields in metals in terms of the magnetic
moment on the FS, and discussed the experimental implications.  Unlike
the CME~\cite{parameswaran-prx14} or the photoinduced
AHE~\cite{mak-science14}, no detailed model of nonequilibrium is
required to quantify the GME, and efficient {\it ab initio} methods
already exist to compute the needed orbital
moments~\cite{lopez-prb12}.

We thank Q.~Niu, J.~Orenstein, D.~Pesin, and D.~Vanderbilt for useful
comments, and also thank D.~Vanderbilt for calling our attention to
Ref.~\cite{yoda-sr15} and suggesting a possible connection with the
present work. We acknowledge support from Grant No.~NSF DMR-1507141
(S.\,Z.), from the DOE LBL Quantum Materials Program and Simons
Foundation (J.\,E.\,M.), and from Grants No.~MAT2012-33720 from the
Spanish Ministerio de Econom\'ia y Competitividad and No.~CIG-303602
from the European Commission (I.\,S.).

{\it Note added.---} Along with the present paper, the role of orbital
moments in the natural gyrotropy of metals was also recognized in
Ref.~\cite{ma-prb15}.

% \bibliography{pap.bib}

%merlin.mbs apsrev4-1.bst 2010-07-25 4.21a (PWD, AO, DPC) hacked
%Control: key (0)
%Control: author (8) initials jnrlst
%Control: editor formatted (1) identically to author
%Control: production of article title (-1) disabled
%Control: page (0) single
%Control: year (1) truncated
%Control: production of eprint (0) enabled
%

\end{document}